\documentclass[11pt,prc,aps,nofootinbib,tightenlines,showpacs,showkeys,preprintnumbers,amsmath,amssymb]{revtex4}

\usepackage{graphicx}
\usepackage{amssymb,amsbsy,amsmath,amsfonts}
\usepackage{hyperref}
\usepackage[english]{babel}

\begin{document}

\title{Structure of $^{14}$C and $^{14}$O nuclei within a
five-cluster model}

\author{B.~E.~Grinyuk}
\author{D.~V.~Piatnytskyi}
\affiliation{Bogolyubov Institute for Theoretical Physics,
National Academy of Sciences of Ukraine, Metrolohichna 14-b, UA-03143 Kyiv, Ukraine}

\begin{abstract}
Within a five-particle model (three $\alpha$-particles plus two
nucleons), the structure functions of mirror nuclei $^{14}$C and
$^{14}$O are studied. Using the variational approach with Gaussian
bases, the energies and wave functions are calculated for these
five-particle systems. Two spatial configurations in the
ground-state wave function are revealed. The r.m.s. charge radius
of $^{14}$O nucleus is found to be $2.415\pm0.005$ fm. The charge
density distributions and the form factors of both nuclei are
predicted. The pair correlation functions are analyzed, and the
r.m.s. relative distances are calculated. The momentum
distributions of particles are found.
\end{abstract}



\date{\today}

\maketitle

\section{Introduction}

Radioactive mirror nuclei $^{14}$C and $^{14}$O in the ground
state are very similar by their structures, a small difference
being produced by an additional Coulomb repulsion in $^{14}$O
nucleus as compared to $^{14}$C one. This fact can be used to
determine the radius of the very rare isotope $^{14}$O not yet
measured experimentally. The idea is to treat both nuclei within
one approach under the condition that the experimental radius of
$^{14}$C is well fitted, as well as the experimental binding
energies of these nuclei. The contribution of the additional
Coulomb repulsion can be taken into account accurately, and thus
the radius of $^{14}$O nucleus can be obtained.

In the present paper, we consider the mirror nuclei $^{14}$C and
$^{14}$O as composed from three $\alpha$-particles and two extra
nucleons (neutrons in $^{14}$C nucleus and protons in $^{14}$O
one). Such model \cite{R1} may have a rather good accuracy as it
was shown by calculations of the structure functions of three- and
four-cluster nuclei \cite{R2,R3,R4,R5,R6} consisting of
$\alpha$-particles and two extra nucleons. For the ground state of
both nuclei and some low-lying energy levels (for which the
excitation of an $\alpha$-particle can be neglected), our
five-particle model can be competitive in accuracy with the
approaches like \cite{R7}, where one has to deal with all the
nucleon degrees of freedom. To solve the five-particle problem, we
exploit the variational method with Gaussian bases \cite{R8,R9}
widely used to study the bound states of few-particle systems.

In the next section, the formulation of the Hamiltonian is given.
The radii of both nuclei and density distributions are
discussed in section 3. In sections 4, 5, and 7, we present the
rest main structure functions of $^{14}$C and $^{14}$O nuclei
(elastic charge form factors, pair correlation functions, and the
momentum distributions of $\alpha$-particles and extra nucleons).
In section 6, we study the characteristic features of the
ground-state wave functions of both nuclei, where two spatial
configurations are revealed.

\section{Statement of the problem}

Within our five-particle model, the Hamiltonian for $^{14}$O
nucleus is assumed to have the form:
\[
\hat{H} = \sum\limits_{i=1}^{2} \frac{\mathbf{p}_{i}^{2}}{2m_{p}}+
\sum\limits_{i=3}^{5}
\frac{\mathbf{p}_{i}^{2}}{2m_{\alpha}}+U_{pp}\left(r_{12}\right)+
\sum\limits_{j>i=3}^{5}\hat{U}_{\alpha\alpha}\left(r_{ij}\right)+
\]
\vskip-2mm
\begin{equation}
+\sum\limits_{i=1}^{2}\sum\limits_{j=3}^{5}
\hat{U}_{p\alpha}\left(r_{ij}\right)+\sum\limits_{j>i=1}^{5}
\frac{Z_{i}Z_{j}e^{2}}{r_{ij}}, \label{E1}
\end{equation}
where indices $p$ and $\alpha$ denote a proton and
an $\alpha$-particle, respectively. In (1), $Z_{1}=Z_{2}=1$ are the
charges of protons (in units of elementary charge $e$), and
$Z_{3}=Z_{4}=Z_{5}=2$ are the charges of $\alpha$-particles (in
the same units).

Within our model, the Hamiltonian for $^{14}$C nucleus is very
similar to (\ref{E1}), but with $Z_{1}=Z_{2}=0$, since the
neutrons have zero charge. We assume the potentials
$\hat{U}_{\alpha\alpha}$ are the same for both nuclei, as well as
the interaction potential between the extra neutrons $U_{nn}$
coincides with $U_{pp}$ due to the charge independence of nuclear
forces. As for the potentials $\hat{U}_{p\alpha}$ in the case of
$^{14}$O nucleus and $\hat{U}_{n\alpha}$ in the case of $^{14}$C
one, their parameters may be slightly different, because the
density distributions of protons and neutrons inside the
$\alpha$-particle are not identical mainly due to the Coulomb
repulsion between protons. We note that the local potential
$U_{nn}$ in the singlet state was successfully used in studying
the structure of $^{6}$He \cite{R4} and $^{10}$Be \cite{R6}
nuclei. The potentials $U_{n\alpha}$ and $U_{p\alpha}$, as well as
the interaction potential between $\alpha$-particles
$U_{\alpha\alpha},$ are of a generalized type with local and
non-local (separable) terms. This type of potentials was first
proposed \cite{R10,R11} to simulate the exchange effects between
particles in interacting clusters and was successfully used, in
particular, in calculations \cite{R2,R4,R6} of multicluster
nuclei. The parameters of the interaction potentials used in this
work are given in \cite{R1}, and the calculations within our model
with these potentials result in the experimental binding energies
of both nuclei under consideration, as well as the experimental
charge radius of $^{14}$C nucleus.

The ground-state energy and the wave function are calculated with
the use of variational method in the Gaussian representation
\cite{R8,R9}, which proved its high accuracy in calculations of
few-particle systems. For the ground $J^{\pi}=0^{+}$ state, the
wave function can be expressed in the form
\begin{equation}
\Phi=\hat{S}\sum_{k=1}^{K}C_{k}\varphi_{k}\equiv\hat{S}
\sum_{k=1}^{K}C_{k}\,\,exp\left(-\sum_{j>i=1}^{5}a_{k,ij}
\left(\mathbf{r}_{i}-\mathbf{r}_{j}\right)^{2}\right), \label{E2}
\end{equation}%
where $\hat{S}$ is the symmetrization operator, and the linear
coefficients $C_{k}$ and nonlinear parameters $a_{k,ij}$ are
variational parameters. The greater the dimension $K$ of the
basis, the more accurate the result can be obtained. The linear
coefficiens can be found within the Galerkin method from the system
of linear equations determining the energy of the system:
\begin{equation}
\sum_{m=1}^{K}C_{m}\left\langle\hat{S}\varphi_{k}
\left|\hat{H}-E\right|\hat{S}\varphi_{m}\right\rangle=0,\,\,\,\,k\,
=\,0,1,...,K. \label{E3}
\end{equation}%
The matrix elements in (\ref{E3}) are known to have explicit form
for potentials like the Coulomb potential or the ones having a
Gaussian expansion. Our potentials \cite{R1} between particles
just have the form of a few Gaussian functions, including the
Gaussian form factor in the separable repulsive term. Thus, system
(\ref{E3}) becomes a system of algebraic equations. We achieved
the necessary high accuracy, by using up to $K=400$ functions of
the Gaussian basis. To fix the non-linear variational parameters
$a_{k,ij}$, we used both the stochastic approach \cite{R8,R9} and
regular variational methods. This enables us to obtain the best
accuracy at reasonable values of the dimension $K$.

As a result, we have the
wave functions of the ground state for both nuclei under consideration
within the five-particle model.
This enables us to analyze the structure functions of $^{14}$C and
$^{14}$O nuclei. In the next section, the charge density
distributions and charge r.m.s. radii are discussed.

\section{Density distributions and radii of $^{14}$C and $^{14}$O nuclei}

The probability density distribution $n_{i}\left(r\right)$ of the
$i$-th particle in a system of particles with the wave function
$|\Phi\rangle$ is known to be
\begin{equation}
n_{i}\left(r\right)=\left\langle\Phi\right|\delta\left(\mathbf{r}-
\left(\mathbf{r}_{i}-\mathbf{R}_{\mathrm{c.m.}}\right)\right)\left|\Phi\right\rangle,
\label{E4}
\end{equation}%
where $\mathbf{R}_{\mathrm{c.m.}}$ is the center of mass of the
system. The probability density distributions are normalized as
$\int n_{i}\left(r\right)d\mathbf{r}=1$.

In Fig.~1, we depict the values $r^{2}n_{p}\left(r\right)$ and
$r^{2}n_{\alpha}\left(r\right)$, respectively, for the density
distributions of extra protons and $\alpha$-particles in $^{14}$O
nucleus. The profiles very close to those shown in Fig.~1 were
obtained for $^{14}$C nucleus in \cite{R1}. It is clearly seen
that extra nucleons in both nuclei move mainly inside the $^{12}$C
cluster formed by $\alpha$-particles. At the same time, the
secondary maximum of curve $1$ at $r\approx 3.4$ fm means that the
extra protons in $^{14}$O nucleus can be found off the $^{12}$C
cluster (with a probability $\simeq 0.16$). Similarly, the extra
neutrons in $^{14}$C nucleus can be found off the $^{12}$C cluster
(with a probability $\simeq 0.14$). Two above-mentioned maxima on
curve $1$ are the consequence of two configurations distinctly
present in the nuclei under consideration (see below).
\begin{figure}%
\includegraphics [width=\textwidth] {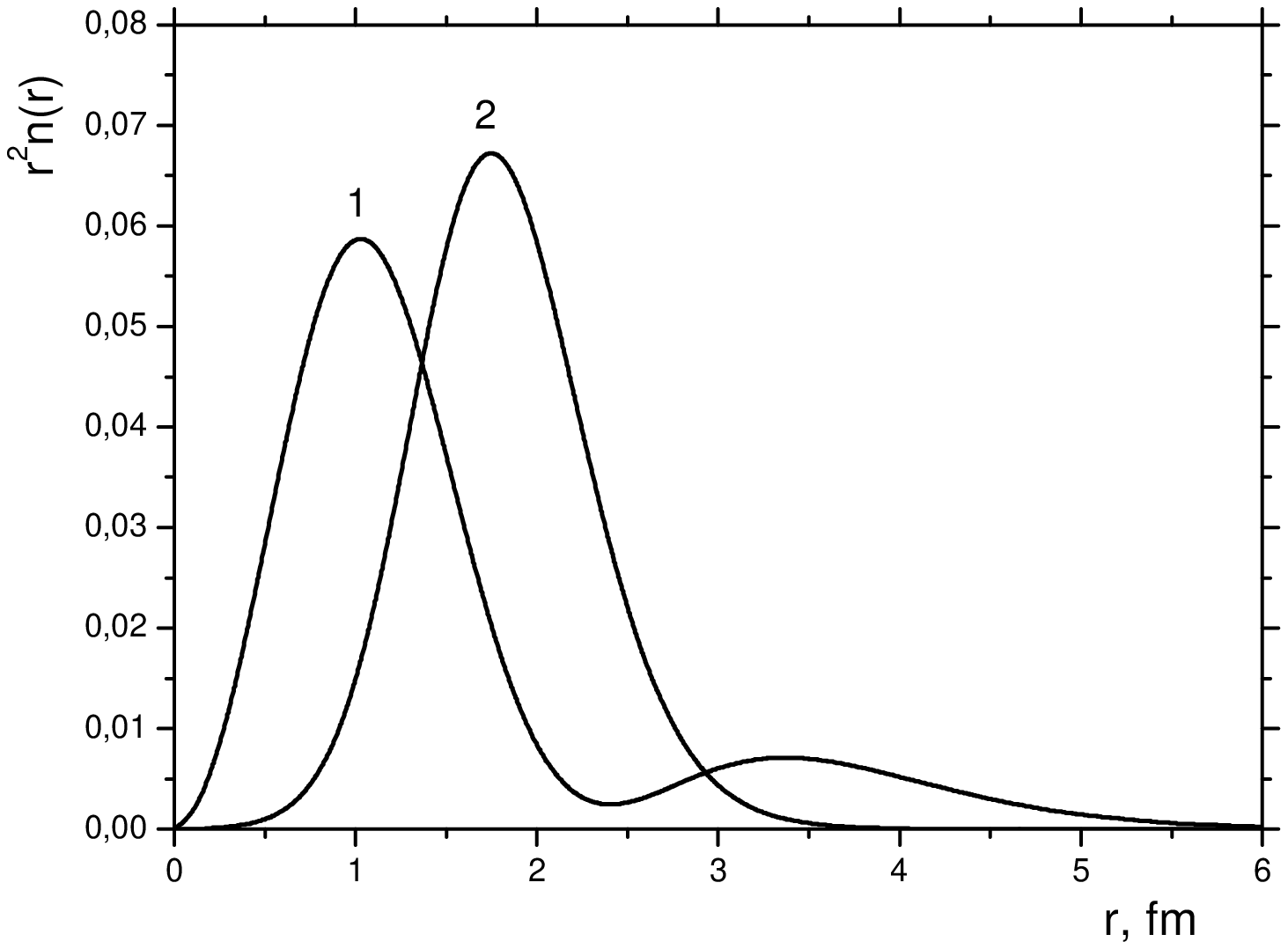}
\vskip-2mm\caption{Probability density distributions multiplied by
$r^{2}$ obtained for extra protons (curve $1$) and
$\alpha$-particles (curve $2$) in $^{14}$O nucleus}
\end{figure}

To find the r.m.s. radius $R_{i}$ of the density distribution
$n_{i}\left(r\right)$, one has to calculate the integral
$R_{i}=\left(\int
r^{2}n_{i}\left(r\right)d\mathbf{r}\right)^{1/2}$. In Table~1, the
main calculated parameters are given for both nuclei. In
particular, the r.m.s. radii $R_{n}$ and $R_{\alpha}$ obtained for
``pointlike'' extra neutrons and $\alpha$-particles in $^{14}$C
nucleus are seen to be less than the corresponding r.m.s. radii
$R_{p}$ and $R_{\alpha}$ for $^{14}$O system. This is explained by
the additional Coulomb repulsion in $^{14}$O nucleus due to the
presence of extra protons instead of extra neutrons present in
$^{14}$C. As a result, the r.m.s. radius of the mass distribution
$R_{m}$ in $^{14}$O is also larger than $R_{m}$ for $^{14}$C. At
the same time, we have $R_{N}<R_{\alpha}$ for both nuclei (where
$R_{N}\equiv R_{n}$ for $^{14}$C and $R_{N}\equiv R_{p}$ for
$^{14}$O). This inequality is directly related to the fact that
the extra nucleons are settled mainly inside the $^{12}$C cluster
(see Fig.~1).

\begin{table*}[!]
\noindent\caption{Calculated energies (MeV) (subtracting the own
energies of $\alpha$-particles), r.m.s. relative
distances, and r.m.s. radii (fm) for \boldmath$^{14}$C and \boldmath$^{14}$O nuclei}
\vskip3mm\tabcolsep6.0pt

\noindent{\footnotesize\begin{tabular}{l c c c c c c c c c }
 \hline \multicolumn{1}{c}
{\rule{0pt}{9pt} Nucleus } &
\multicolumn{1}{|c}{\,\,\,\,\,\,\,\,\,$E$\,\,\,\,\,\,\,\,\,} &
\multicolumn{1}{|c}{\,\,\,\,$r_{NN}$\,\,\,\,} &
\multicolumn{1}{|c}{\,\,\,\,$r_{N\alpha}$\,\,\,\,} &
\multicolumn{1}{|c}{\,\,\,\,$r_{\alpha\alpha}$\,\,\,\,}&
\multicolumn{1}{|c}{\,\,\,\,$R_{N}$\,\,\,\,}&
\multicolumn{1}{|c}{\,\,\,\,$R_ {\alpha}$\,\,\,\,} &
\multicolumn{1}{|c}{\,\,\,\,$R_ {m}$\,\,\,\,} &
\multicolumn{1}{|c}{\,\,\,\,$R_ {\mathrm{ch}}$\,\,\,\,}
\\%
\hline%
\,\,\,\,\,\, $^{14}$C & $-20.398$ & $2.621$ & $2.667$ & $3.189$ &
$1.786$ & $1.852$ & $2.433$ & $2.500$
\\[0.5mm]
 \hline
\,\,\,\,\,\, $^{14}$O & $-13.845$ & $2.732$ & $2.750$ & $3.239$ &
$1.864$ & $1.882$ & $2.461$ & $2.415$
\\[0.5mm]
 \hline
\end{tabular}}
\end{table*}

To find the charge density distributions of the nuclei with regard
for finite sizes of $\alpha$-particles and extra protons, we use
the Helm approximation \cite{R12,R13}. Within this approach, the
charge density distribution for $^{14}$C nucleus,
\begin{equation}
n_{\mathrm{ch}}\left(r\right)= \int
n_{\alpha}\left(\left|\mathbf{r}-\mathbf{r}'\right|\right)
n_{\mathrm{ch,^{4}He}}\left(r'\right) d\mathbf{r}', \label{E5}
\end{equation}%
is a convolution product of the density distribution $n_{\alpha}$ for the probability
to find an $\alpha $-particle inside the $^{14}$C
nucleus with the charge density distribution
$n_{\mathrm{ch,^{4}He}}$ of an $\alpha$-particle itself. The value of
$n_{\alpha}$ is calculated within our five-particle model, while
$n_{\mathrm{ch,^{4}He}}$ follows from the experimental form factor
\cite{R14}. In relation (\ref{E5}), we neglect the small
contribution of extra neutrons. We normalize the charge density
distribution as $\int n_{\mathrm{ch}}\left(r\right)d\mathbf{r}=1$,
i.e. one has to multiply it by $Ze$ to obtain the necessary
dimensional units.

In case of $^{14}$O nucleus, instead of (\ref{E5}), we have two following
terms in the expression for $n_{\mathrm{ch}}\left(r\right)$ due to
the role of extra protons:
\begin{equation}\label{E6}
n_{\mathrm{ch}}\left(r\right)= \frac{3}{4}\int
n_{\alpha}\left(\left|\mathbf{r}-\mathbf{r}'\right|\right)
n_{\mathrm{ch,^{4}He}}\left(r'\right) d\mathbf{r}'+\frac{1}{4}\int
n_{p}\left(\left|\mathbf{r}-\mathbf{r}'\right|\right)
n_{\mathrm{ch},p}\left(r'\right) d\mathbf{r}'
\end{equation}
(the coefficients 3/4 and 1/4 are
proportional to the charges of three $\alpha$-particles and
two extra protons, respectively).

The charge density distributions for $^{14}$C and $^{14}$O nuclei
are shown in Fig.~2. For comparison, the dashed line presents the
density distribution of the probability to find an $\alpha$-particle in
$^{14}$C nucleus (almost the same distribution could be shown for
$^{14}$O). We distinctly see a dip near the origin in the
probability density distribution of ``poinlike'' $\alpha$-particles,
but the integration in the convolution product (\ref{E5}), i.e., the
account for a finite size of $\alpha$-particles, results in the maximum
of the charge density distribution of $^{14}$C nucleus near $r=0$,
nothing to say about $^{14}$O nucleus, where the contribution of
charged extra protons in (\ref{E6}) makes the maximum at $r=0$
even more pronounced. Since the distributions are normalized to 1,
the charge density distribution of $^{14}$O nucleus should be
lower than that of $^{14}$C one at larger distances important for
the calculation of the r.m.s. charge radii. As a result, we have the
inequality:
\begin{equation}
R^{2}_{\mathrm{ch,^{14}O}}\equiv\int r^{2}n_{\mathrm{ch,
^{14}O}}\left(r\right)d\mathbf{r} <
R^{2}_{\mathrm{ch,^{14}C}}\equiv\int r^{2}n_{\mathrm{ch,
^{14}C}}\left(r\right)d\mathbf{r}.
\end{equation}\label{E7}
Within the Helm approximation for the density distributions
(\ref{E5}) and (\ref{E6}), the charge radii squared are known to be
\begin{equation}\label{E8}
R^{2}_{\mathrm{ch,^{14}C}}=R^{2}_{\mathrm{\alpha,^{14}C}}+R^{2}_{\mathrm{ch,^{4}He}},
\end{equation}
\begin{equation}\label{E9}
R^{2}_{\mathrm{ch,^{14}O}}=\frac{3}{4}\left(R^{2}_{\mathrm{\alpha,^{14}O}}+R^{2}_{\mathrm{ch,^{4}He}}\right)+
\frac{1}{4}\left(R^{2}_{\mathrm{p,^{14}O}}+R^{2}_{\mathrm{ch,p}}\right).
\end{equation}
Due to the fact that $R_{\mathrm{p,^{14}O}} <
R_{\mathrm{\alpha,^{14}O}}$, i.e. the extra protons move (on the
average) closer to the center of $^{14}$O nucleus than the
$\alpha$-particles do (see Table 1), and due to the well-known
experimental fact that $R_{\mathrm{ch,p}} <
R_{\mathrm{ch,^{4}He}}$, we again confirm inequality (\ref{E7}).
The calculated r.m.s. charge radii are given in Table~1. Since the
errors of the experimental radius of an $\alpha$-particle
\cite{R15,R16} are of the order of $\pm 0.003$ fm, we estimate the
accuracy of our result for the charge r.m.s. radius of $^{14}$O to
be of the order of $\pm 0.005$ fm. We hope for that the future
experiments should confirm the calculated r.m.s. charge radius of
$^{14}$O nucleus: $R_{\mathrm{ch,^{14}O}} = 2.415\pm 0.005$ fm.

It is obvious that the mass r.m.s. radius of $^{14}$O nucleus is,
vice versa, greater than that of $^{14}$C nucleus (see Table~1),
because both nuclei have very similar mass distribution structures,
and $^{14}$O nucleus has the lower binding energy and larger relative
distances between particles due to the additional Coulomb
repulsion of extra protons.
\begin{figure}%
\includegraphics [width=\textwidth] {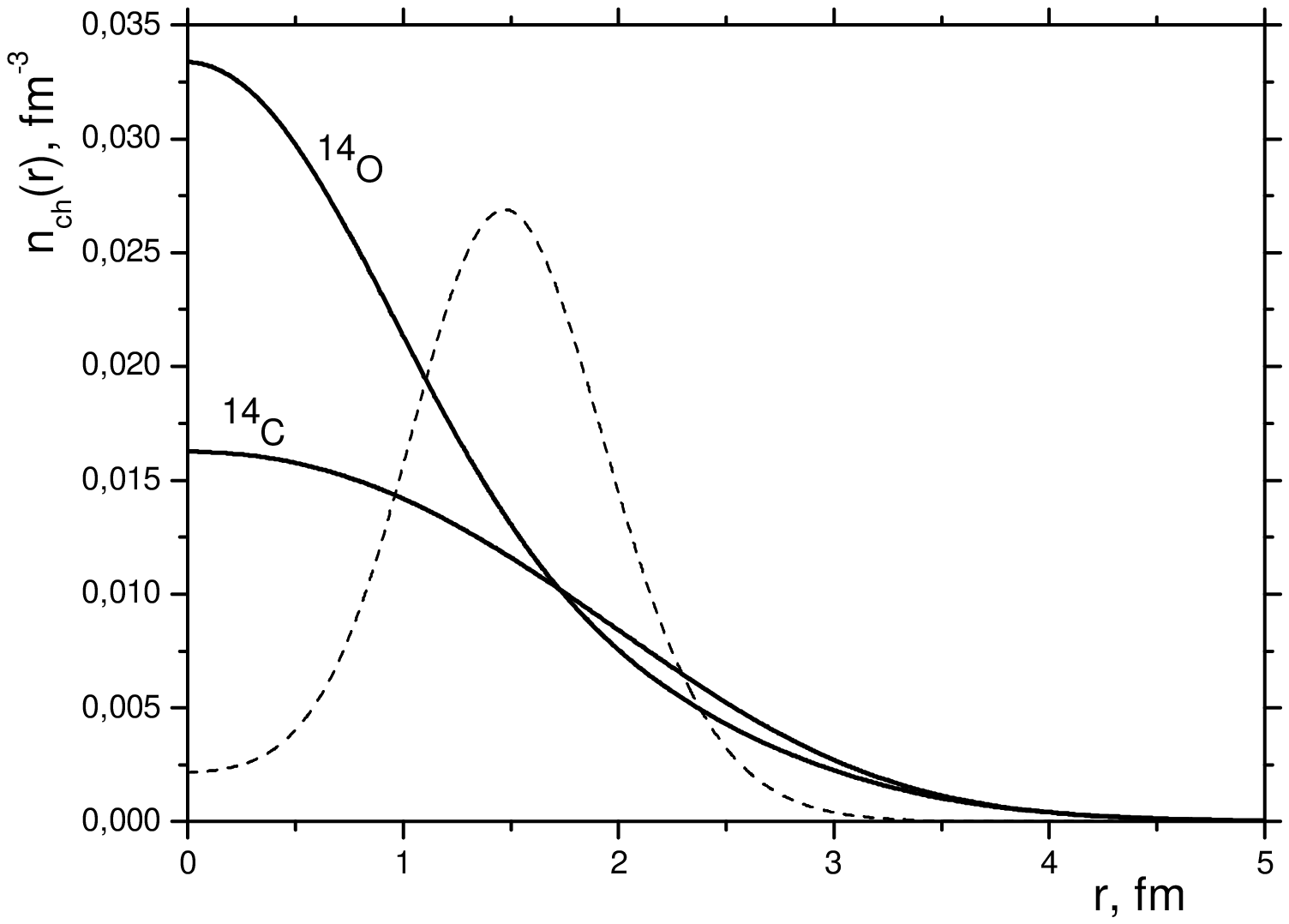}
\vskip-2mm\caption{Charge density distributions in $^{14}$C and
$^{14}$O nuclei (normalized as $\int
n_{ch}\left(r\right)d\mathbf{r}=1$). The dashed line depicts the
probability density distribution of $\alpha$-particles in $^{14}$C
nucleus}
\end{figure}

\section{Elastic charge form factors of $^{14}$C and $^{14}$O nuclei}

The elastic charge form factor $F_{\mathrm{ch}}\left(q\right)$ can
be found by a Fourier transformation of the density distribution
$n_{\mathrm{ch}}\left(r\right)$,
\begin{equation}\label{E10}
F_{\mathrm{ch}}\left(q\right)=\int \exp
\left(-i\left(\mathbf{q}\mathbf{r} \right)\right)
n_{\mathrm{ch}}\left(r\right)d\mathbf{r}.
\end{equation}
In the Helm approximation for $n_{\mathrm{ch}}\left(r\right)$ of
$^{14}$C nucleus, the convolution product (\ref{E5}) for the
charge density distribution transforms to a product, and for the
charge form factor of $^{14}$C one has
\begin{equation}\label{E11}
F_{\mathrm{ch, ^{14}C}}\left(q\right)=F_{\mathrm{\alpha,^{14}C}}
\left(q\right)F_{\mathrm{ch,^{4}He}} \left(q\right),
\end{equation}
where $F_{\mathrm{\alpha,^{14}C}}\left(q\right)$ is the Fourier
transform of the probability density distribution of
$\alpha$-particles in $^{14}$C nucleus, and
$F_{\mathrm{ch,^{4}He}} \left(q\right)$ is the form factor of
$^{4}$He. The profile of
$F_{\mathrm{\alpha,^{14}C}}\left(q\right)$ is calculated within
our five-particle model, and the form factor of $^{4}$He nucleus
is taken from the experiment \cite{R14}.

It is clear from expression (\ref{E11}) that the elastic charge
form factor of $^{14}$C nucleus should become zero at that
momentum transfer squared $q^{2},$ where any of two multipliers in
(\ref{E11}) becomes zero. The absolute value of the form factor at
corresponding values of $q^{2}$ should have ``dips''. In Fig.~3,
we depict the absolute value of the elastic charge form factor
(\ref{E11}) (solid line) together with both multipliers from the
right-hand side of relation (\ref{E11}) (dashed lines). The dip at
$q^{2}\simeq 10$ fm$^{-2}$ is intrinsic to $^{4}$He nucleus
\cite{R14}, and it reveals itself in all the form factors of light
nuclei obtained within the $\alpha$-particles plus some neutrons
model (see, e.g., form factors of $^6$He \cite{R4} or $^{10}$Be
\cite{R6}). The dip at $q^{2}\simeq 3.7$ fm$^{-2}$ is proper to
the form factor $F_{\mathrm{\alpha,^{14}C}} \left(q\right)$ in
(\ref{E11}). Since its properties are determined by three
$\alpha$-particles forming the $^{12}$C cluster inside the
$^{14}$C nucleus, it is not surprising that we observe the first
dip for the form factor of $^{12}$C nucleus \cite{R17,R18} at
almost the same $q^{2}\simeq 3.4$ fm$^{-2}$. This means that the
$^{12}$C cluster in $^{14}$C nucleus is only slightly disturbed by
two extra neutrons, and just this cluster is responsible for the
first dip in the elastic charge form factor of $^{14}$C nucleus.

Passing to the charge form factor of $^{14}$O nucleus, we
should consider the contribution of extra protons to the
charge density distribution (\ref{E6}) of this system. As a
result, the elastic form factor, i.e. the Fourier transform
of the charge density distribution (\ref{E6}), contains two
terms:
\begin{equation}\label{E12}
F_{\mathrm{ch, ^{14}O}}\left(q\right)=
\frac{3}{4}F_{\mathrm{\alpha,^{14}O}}
\left(q\right)F_{\mathrm{ch,^{4}He}}\left(q\right)+\frac{1}{4}
F_{p\mathrm{,^{14}O}}\left(q\right)F_{\mathrm{ch,}p}
\left(q\right),
\end{equation}
where $F_{p\mathrm{,^{14}O}}\left(q\right)$ is a Fourier
transform of the probability density distribution of extra
protons in $^{14}$O nucleus, and $F_{\mathrm{ch,}p}\left(q\right)$
is the charge form factor of the proton itself \cite{R19}. The
rest notations are similar to ones in formula (\ref{E11}).
Although the first term on the right-hand side of (\ref{E12}) is very
similar to expression (\ref{E11}) and becomes zero at the same
momentum transfer squared $q^{2}$ as the form factor (\ref{E11}) does,
the second term in (\ref{E12}) (present due the extra protons)
is a smoothly decreasing function. Thus, the total expression
does not become zero, though the first term does. In Fig.~4, the
absolute value of the elastic charge form factor of $^{14}$O nucleus
is shown (solid line 1). In the contrary to the form factor of
$^{14}$C nucleus (Fig.~3), it has no dips within the presented
range of $q^{2}$. The dashed lines show the calculated
$F_{p\mathrm{,^{14}O}}\left(q\right)$ (dashed line 2) and
the experimental form factor $F_{\mathrm{ch,}p}\left(q\right)$
\cite{R19} (dashed line 3).
\begin{figure}%
\includegraphics [width=\textwidth] {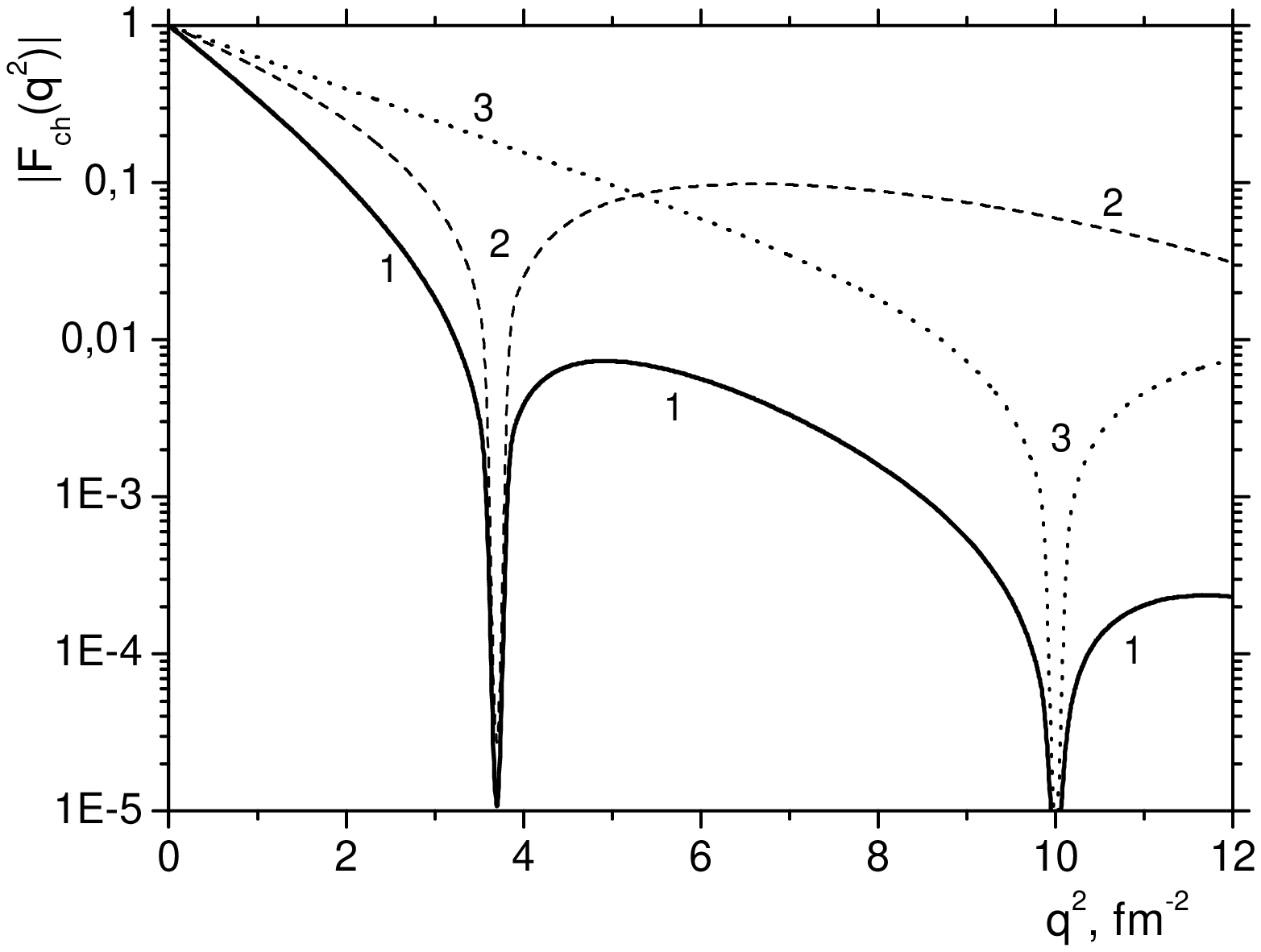}
\vskip-2mm\caption{Elastic charge form factor of $^{14}$C nucleus
(solid line 1). Dashed line 2 depicts the calculated form
factor $F_{\mathrm{\alpha,^{14}C}} \left(q\right)$, and dashed
line 3 depicts the experimental form factor \cite{R14} of $^{4}$He
nucleus}
\end{figure}

\begin{figure}%
\includegraphics [width=\textwidth] {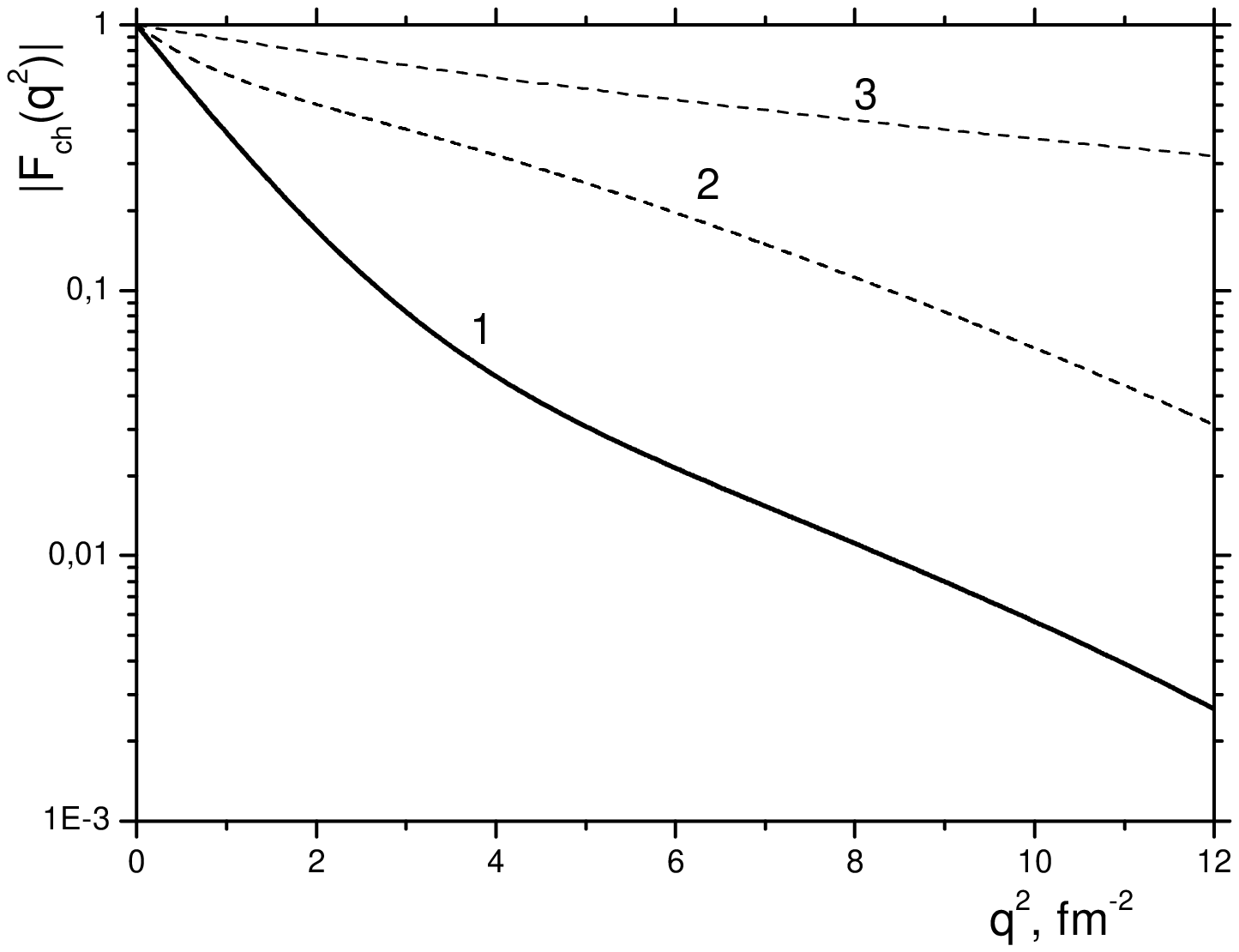}
\vskip-2mm\caption{Elastic charge form factor of $^{14}$O nucleus
(solid line 1). Dashed lines 2 and 3 present the calculated
form factor $F_{\mathrm{p,^{14}O}} \left(q\right)$ and
experimental form factor
$F_{\mathrm{ch,}p}\left(q\right)$ of a proton \cite{R19}, respectively}
\end{figure}

\section{Pair correlation functions and relative distances}

More information about the structures of $^{14}$C and $^{14}$O
nuclei can be obtained from the analysis of the pair correlation
functions. The pair correlation function $g_{ij}\left(r\right)$
for a pair of particles $i$ and $j$ can be defined as follows:
\begin{equation}\label{E13}
g_{ij}\left(r\right)=\left\langle\Phi\right|\delta\left(\mathbf{r}-
\left(\mathbf{r}_{i}-\mathbf{r}_{j}\right)\right)
\left|\Phi\right\rangle,
\end{equation}
and it is known to be the density of the probability to find the
particles $i$ and $j$ at a definite distance $r$. The r.m.s.
relative distances squared $\left\langle r^{2}_{ij}\right\rangle$
are directly expressed through the pair correlation functions
$g_{ij}$:
\begin{equation}\label{E14}
\left\langle r^{2}_{ij}\right\rangle=\int r^{2}
g_{ij}\left(r\right)d \mathbf{r}.
\end{equation}
The calculated r.m.s. relative distances between particles are
given in Table~1 for $^{14}$C and $^{14}$O nuclei. As was mentioned
above, due to the additional Coulomb repulsion in $^{14}$O as compared
to $^{14}$C nucleus, all the corresponding relative distances in
$^{14}$O nucleus are greater than those in $^{14}$C. We note that
the r.m.s. radii $R_{i}$ are connected with the r.m.s. relative
distances $r_{jk}$:
\begin{equation}\label{E15}
R_{i}^{2}=\frac{1}{M^{2}}\left(\left(M-m_{i}\right)\sum_{j\neq i}
m_{j}r_{ij}^{2}\,\,\, -\sum_{j<k \, \left(j\neq i, k\neq
i\right)}m_{j}m_{k}r_{jk}^{2}\right),
\end{equation}
where $M$ is the total mass of the system of particles. Thus,
the r.m.s. radii could be calculated with the use of the pair
correlation functions and relations (\ref{E14}) and (\ref{E15}).

Since the average of a pairwise local potential
$V_{ij}\left(r\right)$ is expressible directly through the pair
correlation function $g_{ij}\left(r\right)$,
\begin{equation}\label{E16}
\left\langle \Phi \right|V_{ij}\left|\Phi\right\rangle=\int
V_{ij}\left(r\right)g_{ij}\left(r\right)d\mathbf{r},
\end{equation}
the variational principle makes the profile of
$g_{ij}\left(r\right)$ such that it has a maximum, where the
potential is attractive, and a minimum in the area of repulsion (if
the role of the kinetic energy is not crucial). The
$\alpha$-particles have about four times greater mass than extra
nucleons, and thus their kinetic energy is essentially smaller
than that of nucleons (see below). As a result, the pair
correlation function $g_{\alpha \alpha}\left(r\right)$ profile is
determined mainly by the potential $\hat{U}_{\alpha \alpha}$ and
has a pronounced maximum (curve 1 in Fig.~5) near the minimum of
the potential attraction. On the other hand, due to the presence of a
local repulsion in the same potential near the origin, the profile
of $g_{\alpha \alpha}\left(r\right)$ has a dip at short distances.
Thus, the profile of $g_{\alpha \alpha}\left(r\right)$ shows that
$\alpha$-particles are mainly settled at a definite distances
$r_{\alpha\alpha}$ (see Table~1) and form a triangle of $^{12}$C
cluster.

The extra nucleon pair correlation function
($g_{nn}\left(r\right)$ for $^{14}$C, as well as
$g_{pp}\left(r\right)$ for $^{14}$O), also has a dip at short
distances (see Fig.~5, curve 2) due to the presence of a
short-range repulsion in our singlet nucleon-nucleon potential
\cite{R1,R4,R6}. The function $g_{n\alpha}\left(r\right)$ has no
pronounced dip in the origin, since our model \cite{R1} of
generalized potential between a nucleon and an $\alpha$-particle
contains the local pure attractive potential plus the non-local
(separable) repulsion of greater radius.

\begin{figure}%
\includegraphics [width=\textwidth] {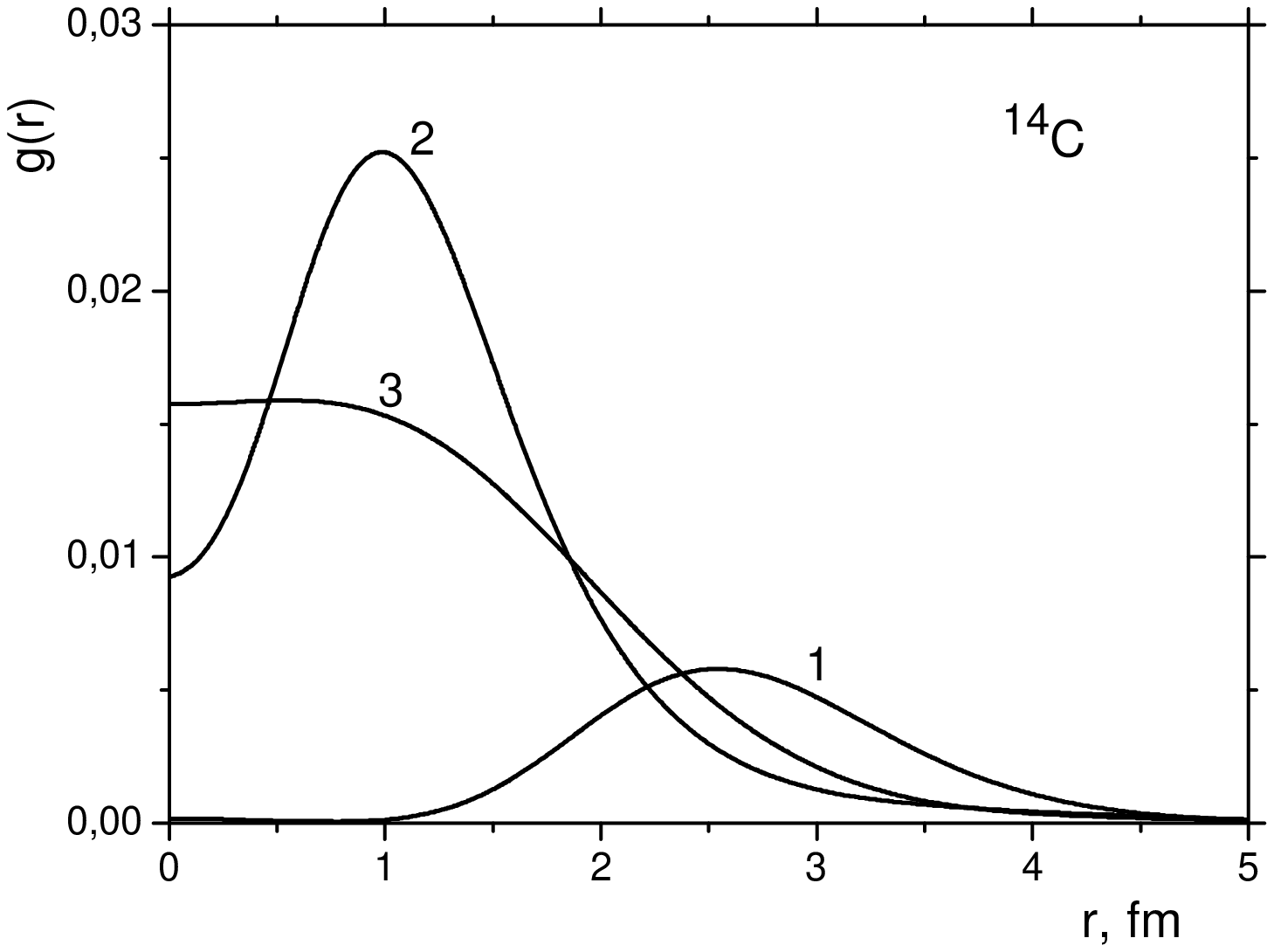}
\vskip-2mm\caption{Pair correlation functions for $^{14}$C
nucleus: $g_{\alpha\alpha}(r)$ - curve 1, $g_{nn }(r)$ - curve 2,
and $g_{n\alpha}(r)$ - curve 3}
\end{figure}

\section{Two configurations in $^{14}$C and $^{14}$O nuclei}

To make the structure of the ground state of $^{14}$C and $^{14}$O
nuclei more clear, let us consider the quantity
$P\left(r,\rho,\theta\right)$ proportional to the density of the
probability to find extra nucleons at a definite relative distance
$r$ and to find their center of mass at a distance $\rho$ from the
center of mass of $^{12}$C cluster:
\begin{equation}\label{E17}
P\left(r,\rho,\theta\right)=r^{2}\rho^{2}\left\langle\Phi\right|\delta\left(\mathbf{r}-\mathbf{r}_{NN}\right)
\delta\left(\boldsymbol{\rho}-\boldsymbol{\rho}_{(NN),(3\alpha)}\right)\left|\Phi\right\rangle,
\end{equation}
where $\theta$ is the angle between the vectors $\mathbf{r}$ and
$\boldsymbol{\rho}$. The value $P\left(r,\rho,\theta\right)$ is
depicted in Fig.~6 for $\theta=0^{\circ}$, $\theta=30^{\circ}$,
$\theta=45^{\circ}$, and $\theta=90^{\circ}$ as a function of $r$
and $\rho$. Two peaks on the $P\left(r,\rho,\theta\right)$ surface
are observed at $\theta=0^{\circ}$, and only one peak at
$\theta=90^{\circ}.$ The rest angles give intermediate results
(see Fig.~6 for $\theta=30^{\circ}$ and $\theta=45^{\circ}$). If
it were not the multiplier $r^{2}\rho^{2}$ in (\ref{E17}), the
main peak present at all the angles $\theta$ would be settled just
at $\rho=0$, i.e. the center of mass of $^{12}$C cluster and that
of the dinucleon one would coincide.  The second peak reveals
itself mainly at $\theta=0^{\circ}$, and it corresponds to a
configuration, where the dinucleon subsystem touches the center of
$^{12}$C cluster by one of the extra nucleons, and another one is
comparatively far from the center of the nucleus (it is out of
$^{12}$C cluster). Just this configuration makes a contribution to
the second maximum of the extra nucleon probability density
distribution (see Fig.~1). In this configuration, the center of
mass of the subsystem of extra nucleons does not coincide with the
center of mass of $^{12}$C cluster. A similar situation with two
configurations in the ground state is found for $^{6}$He, $^{6}$Li
\cite{R2,R3,R4,R7} or $^{10}$Be, $^{10}$C \cite{R5,R6} nuclei,
where the center of mass of the dinucleon subsystem coincides (one
configuration) or does not coincide (another configuration) with
the center of mass of the subsystem of $\alpha$-particles.
\begin{figure}%
\includegraphics [width=\textwidth] {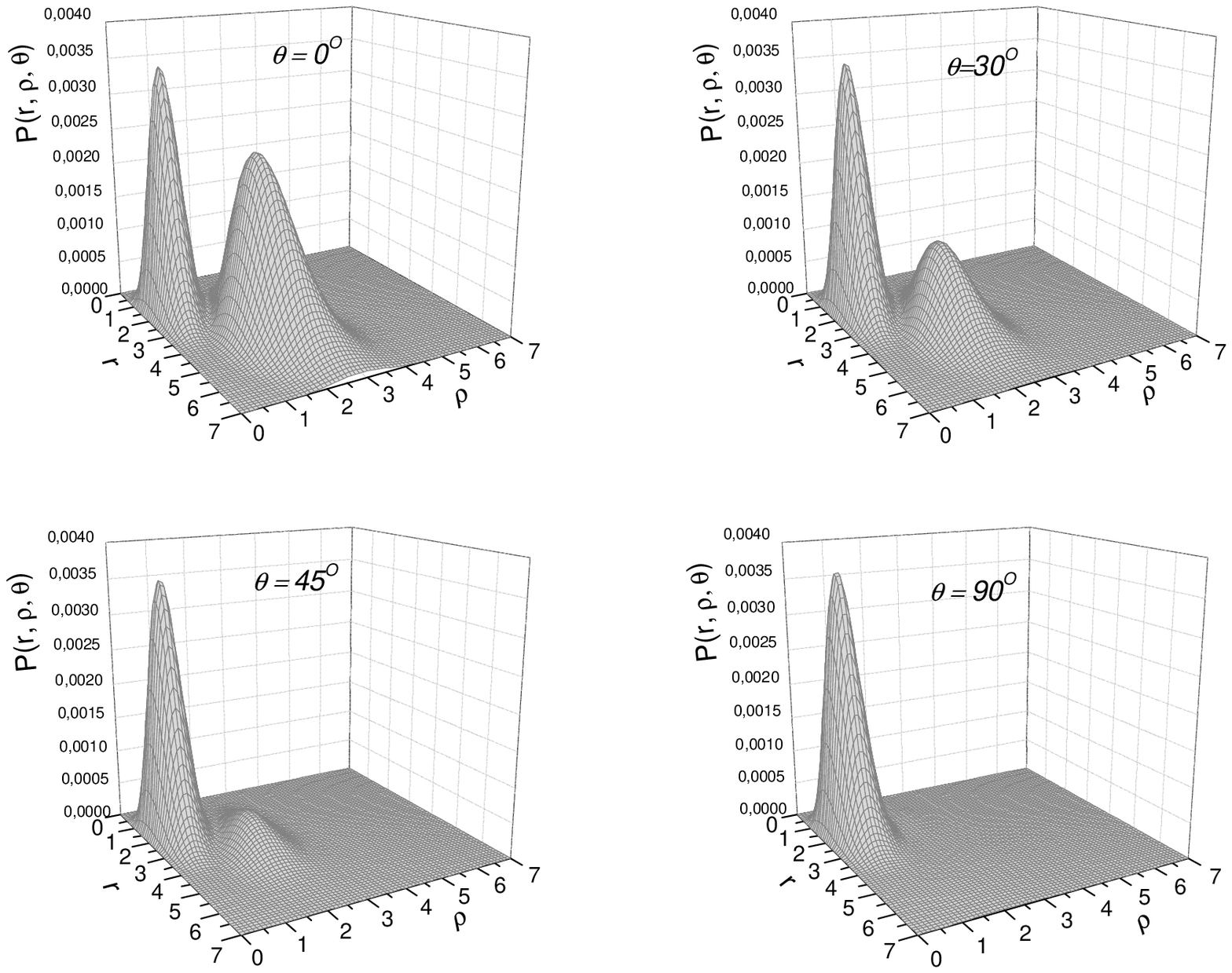}
\vskip-2mm\caption{Two configurations in the ground state of
$^{14}$C nucleus manifesting themselves in the
$P\left(r,\rho,\theta\right)$ function at different angles
$\theta$}
\end{figure}

\section{Momentum distributions}

To complete the study of the structure functions of $^{14}$C and
$^{14}$O nuclei, we present the momentum distributions of
$\alpha$-particles and extra nucleons in these systems within the
five-particle model. The momentum distribution
$n_{i}\left(k\right)$ of the $i$-th particle is known to be the
density of the probability to find this particle with a definite momentum
$k$,
\begin{equation}\label{E18}
n_{i}\left(k\right)=\left\langle\tilde{\Phi}\right|\delta
\left(\mathbf{k}-\left(\mathbf{k}_{i}-\mathbf{K}_{c.m.}\right)\right)
\left|\tilde{\Phi}\right\rangle,
\end{equation}
where $\tilde{\Phi}$ is the wave function of the system in the
momentum representation. The normalization of the momentum
distribution is $\int n_{i}\left(k\right)d\mathbf{k}=1$. The
momentum distribution $n_{i}\left(k\right)$ enables one, in
particular, to calculate the average kinetic energy of the $i$-th
particle:
\begin{equation}\label{E19}
\left\langle
E_{i,kin}\right\rangle=\int\frac{k^{2}}{2m_{i}}n_{i}\left(k\right)d\mathbf{k}.
\end{equation}
Mainly due to the mass ratio between a nucleon and an
$\alpha$-particle, the extra nucleons move much more rapidly than
the $\alpha$-particles inside $^{14}$C and $^{14}$O nuclei. In
particular, the calculated kinetic energy of each of the extra
neutrons in $^{14}$C nucleus is about $32.66$ MeV, while the same
value for an $\alpha$-particle amounts about $6.83$ MeV. For
$^{14}$O nucleus, we have $31.77$ MeV for an extra proton, and
$6.62$ MeV for an $\alpha$-particle. The corresponding
ratio of velocities is about $4.4$ (for both nuclei). This means that the extra
nucleons move essentially faster than the heavier
$\alpha$-particles do.

The momentum distributions are very close for both considered
nuclei. That is why, we present the profiles of the momentum
distributions only for $^{14}$C. In Fig.~7, curve 1 corresponds to
the momentum distribution of an $\alpha$-particle, and curve 2
depicts $n_{n}\left(k\right)$ of an extra neutron.
\begin{figure}%
\includegraphics [width=\textwidth] {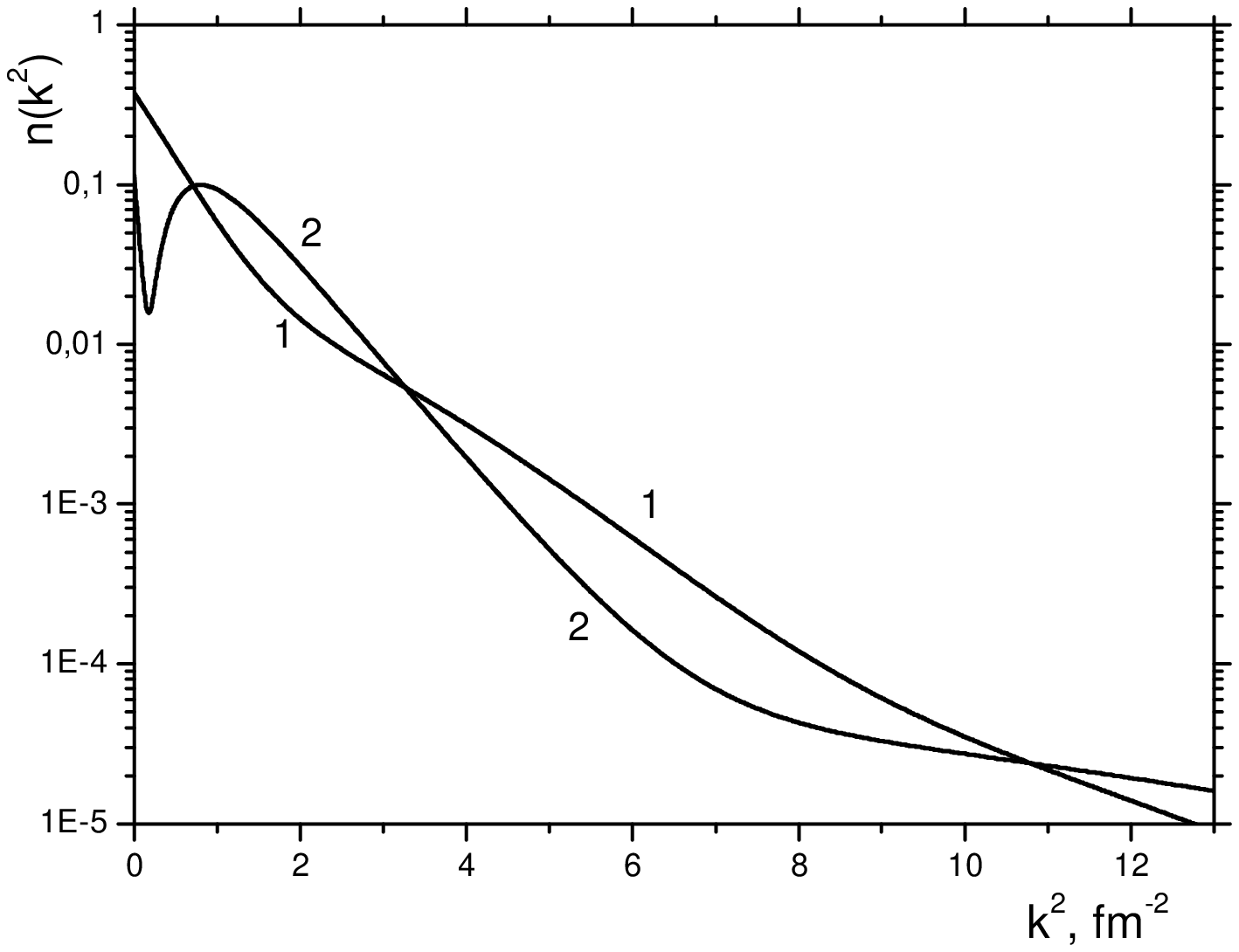}
\vskip-2mm\caption{Momentum distributions of an
$\alpha$-particle (curve 1) and an extra neutron (curve 2) in
$^{14}$C nucleus}
\end{figure}
The momentum distribution of $\alpha$-particles
$n_{\alpha}\left(k\right)$ is seen to be a monotonically
decreasing function, while $n_{n}\left(k\right)$ has two maxima:
at zero momentum and at $k^{2}\simeq 1$ fm$^{-2}$. These two
maxima correspond to two above-mentioned configurations in the
ground state of the nucleus. In a configuration, where an extra
neutron is comparatively far from the center of the nucleus, it
moves comparatively slowly and makes a contribution to the peak at
very small $k^{2}$. As for configurations, where the extra
neutrons move mainly inside $^{12}$C cluster, their momenta are
somewhat greater, and they make their contribution to the second
maximum at $k^{2}\simeq 1$ fm$^{-2}$. At the same time, the
$\alpha$-particles inside $^{12}$C cluster almost do not feel the
extra neutrons motion peculiarities, and, thus, their influence on
the momentum distribution of $\alpha$-particles is small due to
the mass ratio and the comparatively large binding energy of
$^{12}$C cluster.

\section{Conclusions}

To sum up, we note that the mirror nuclei $^{14}$C and $^{14}$O
have very close structures of their ground-state wave functions,
$^{14}$O nucleus having a little bit greater size and a less
binding energy because of the additional Coulomb repulsion due to
the charges of extra protons. At the same time, it is shown that
the r.m.s. charge radius of $^{14}$O nucleus is less than that of
$^{14}$C due to the position of extra nucleons mainly inside
$^{12}$C cluster. The r.m.s. charge radius of $^{14}$O nucleus is
predicted. The charge density distribution and the elastic charge
form factor of $^{14}$C nucleus are shown to be essentially
different from the same values for $^{14}$O, whereas the
distributions independent of the charges of particles are almost
coincide (including probability density distributions of
particles, pair correlation functions, and the momentum
distributions). Two configurations in the ground-state wave
function are revealed, where $^{12}$C cluster and the dinucleon
subsystem have the same centers of mass (first configuration, with
a dinucleon inside $^{12}$C cluster), or shifted centers of mass
(second configuration, with one nucleon outside of $^{12}$C
cluster).

\section*{Acknowledgment}

This work was supported in part by the Program of Fundamental
Research of the Department of Physics and Astronomy of the
National Academy of Sciences of Ukraine (project No. 0112U000056).


\bigskip

\bibliographystyle{elsarticle-num}

\end{document}